\documentclass[superscriptaddress,reprint,10pt,a4paper,onecolumn,nolongbibliography]{revtex4-2}
\usepackage{bm}

\usepackage{hyperref}
\usepackage{amsmath}
\usepackage{siunitx}
\DeclareSIUnit{\angstrom}{\textup{\r A}}
\let\ts\textsubscript
\usepackage{graphicx}
\usepackage{xcolor}
\usepackage{makecell}
\usepackage[normalem]{ulem}
\usepackage{enumitem}
\usepackage{algorithm}
\usepackage{algpseudocode}

\newcommand{\textsub}[2]{{#1}_{\text{#2}}}
\newcommand{\ii}{\mathrm{i}}
\newcommand{\dd}{\mathrm{d}}
\newcommand{\ee}{\mathrm{e}}

\DeclareMathOperator{\svMap}{svMap}
\newcommand{\EQ}[1]{Eq.~(\ref{#1})}
\newcommand{\FIG}[1]{Fig.~\ref{#1}}
\newcommand{\FIGs}[1]{Figs.~\ref{#1}}
\newcommand{\TBL}[1]{Table~\ref{#1}}
\newcommand{\ALGO}[1]{Algorithm~\ref{#1}}

\AtBeginDocument{

\let\Im\relax

\DeclareMathOperator{\Im}{Im}
}

\begin{document}

\title{An atom-based machine-learned dipole-moment model and application to conjugated systems}

\author{Yuansheng Zhao}
\email{yszhao@g.ecc.u-tokyo.ac.jp}
\affiliation{Quemix Inc., Chuo-ku, Tokyo 103-0027, Japan}
\affiliation{Department of Physics, The University of Tokyo, Bunkyo-ku, Tokyo 113-0033, Japan}
\author{Tamio Yamazaki}
\affiliation{JSR--UTokyo Collaboration Hub, CURIE, JSR Corporation, Minato-ku, Tokyo 105-8640, Japan}
\author{Ryohei Hosoya}
\affiliation{Quemix Inc., Chuo-ku, Tokyo 103-0027, Japan}
\affiliation{Department of Physics, The University of Tokyo, Bunkyo-ku, Tokyo 113-0033, Japan}
\author{Yu-ichiro Matsushita}
\affiliation{Quemix Inc., Chuo-ku, Tokyo 103-0027, Japan}
\affiliation{Department of Physics, The University of Tokyo, Bunkyo-ku, Tokyo 113-0033, Japan}
\affiliation{Quantum Materials and Applications Research Center, National Institutes for Quantum Science and Technology (QST), Meguro-ku, Tokyo, 152-8550, Japan}
\author{Shinji Tsuneyuki}
\affiliation{Department of Physics, The University of Tokyo, Bunkyo-ku, Tokyo 113-0033, Japan}
\affiliation{Institute for Physics of Intelligence, The University of Tokyo, Bunkyo-ku, Tokyo 113-0033, Japan}
\affiliation{RIKEN Center for Computational Science (R-CCS), Wako, Saitama 351-0198, Japan}

\begin{abstract}
We introduce a method for predicting the dipole moments of molecular systems 
by systematically decomposing the total dipole moment into the sum of atomic contributions using the wavefunction from density-functional-theory calculations,
and then use graph neutral networks to predict this effective atomic dipole moment from input atomic structure.
It is demonstrated that the dipole moments and dielectric function can be accurately predicted even for complicated conjugated systems where the previous bond-based model [Phys.~Rev.~B \textbf{110}, 165159] fails.
The inference cost of our model scales linearly with the number of atoms and
is about 3 times faster compared with Born-effective-charge-based schemes, but shows similar or better accuracy for dielectric function at terahertz range.
\end{abstract}

\maketitle

\section{Introduction}

The complex dielectric function $\varepsilon(\omega)$ is one of the most fundamental and important material properties that describe the response to an external electric field.
At terahertz (THz) region, it reflects the collective motion of molecules and is a key quantity to study the liquids and biomolecules \cite{siegel2004terahertz} and to design material towards the realization of 6G communication which requires extremely low loss at this frequency range \cite{kakutani2021material,zhai2022terahertz}. 
It is therefore of interest to be able to accurately and efficiently predict the complex dielectric function from computer simulations.

Because the $\varepsilon(\omega)$ can be accessed from the dipole auto-correlation function \cite{kubo1957statistical,neumann1983calculation},
all the required ingredients are simply a trajectory from molecular dynamics (MD) simulation and the dipole moments along the trajectory.
For the former, while relatively short trajectories may be sufficient for infrared (IR) region, to reach THz range, very long trajectories spanning \si{ns} order is required, making accurate \textit{ab initio} MD practically impossible.
This difficulty may be ameliorated thanks to recent development of machine-learned (ML) interatomic potentials \cite{deng2023chgnet,fu2025learning} which provide acceleration by several orders of magnitudes yet still maintains a high accuracy compared with density-functional theory (DFT) \cite{hohenberg1964inhomogeneous,kohn1965self}.
However, the second ingredient, to compute the total dipole moment $\bm M$ of given structures, is also not straightforward.
Classically, this is accomplished by assigning empirical charges $q_I$ at each atom located at $\bm R_I$ and then $\bm M=\sum_Iq_I\bm R_I$.
However, THz dielectric properties are related to collected motion of molecules, and the many-body polarization effects are important \cite{bone2024new,nymand2001temperature}, for non-polar molecules like hydrocarbons in particular, making it challenging for classical models to reach high accuracy in complicated systems. Moreover, assigning non-integral charges may break the quantization condition of dipole moment in periodic systems.

\begin{figure}
\includegraphics[scale=.6]{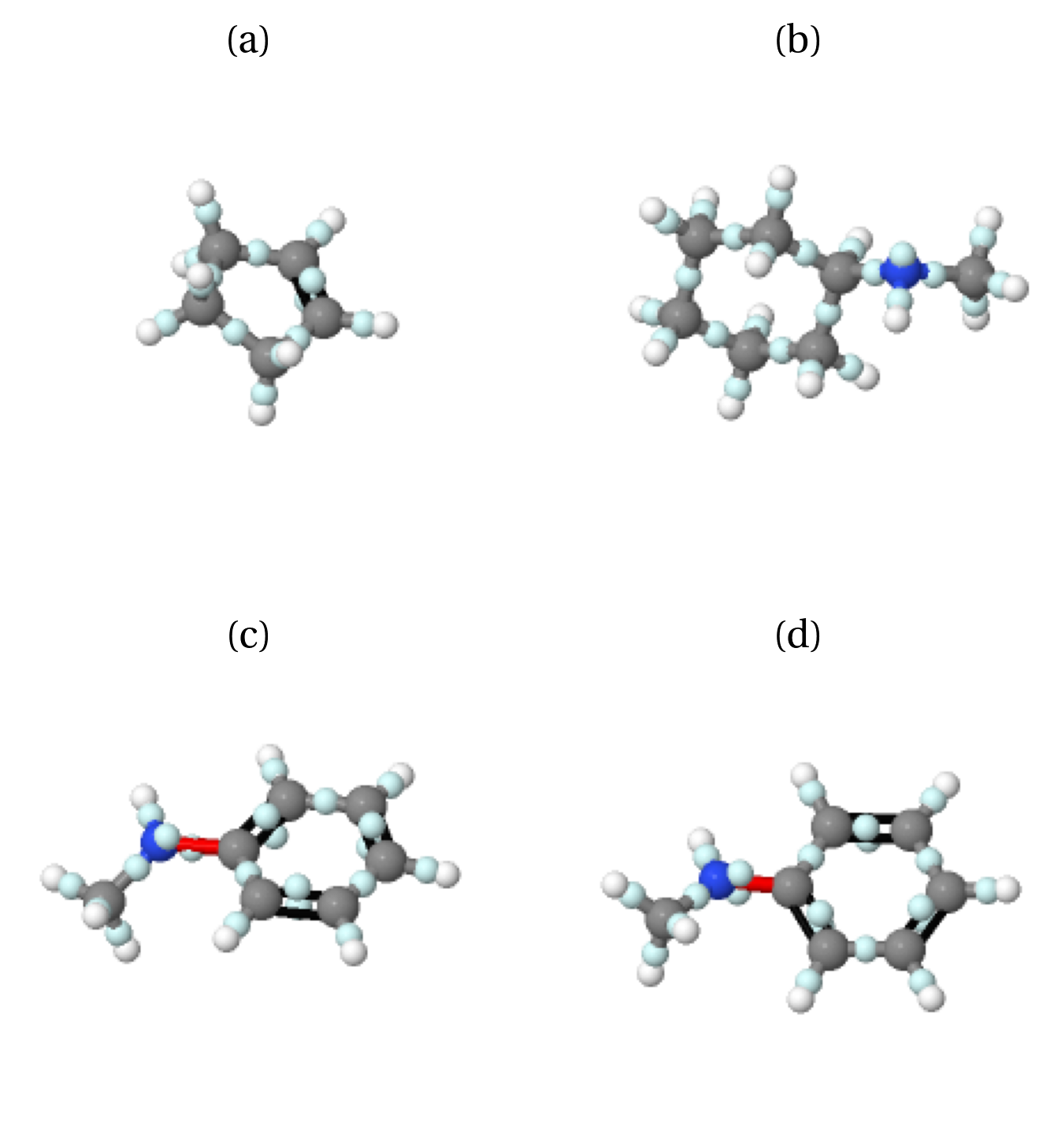}
\caption{Center of maximally localized Wannier functions of several molecules computed with density-functional theory (DFT). Gray, white, blue and light blue balls representing C, H, N atoms and Wannier centers, respectively. (a) Cyclopentene. The C=C double bond is highlighted in black. (b) N-methylcyclohexylamine. The N atom has a lone pair.
(c)(d) Two N-methylaniline molecules. The double bond in benzene ring as indicated by Wannier center is highlighted in black and the red C--N bond also has partial double bond characters.
}\label{fg:mlwc}
\end{figure}

From the modern theory of polarization, the dipole moment can be evaluated by placing charge $-2e$ at the center of Wannier functions \cite{marzari2012maximally} which can be obtained from \textit{ab initio} calculations such DFT.
Therefore, if the location of Wannier centers can be efficiently predicted for given atomic structures, the dipole moment with first-principle accuracy can then be cheaply computed \cite{zhang2020deep,krishnamoorthy2021dielectric}.
Based on the fact that for molecular systems, the Wannier centers are generally located near the bond centers or lone pairs as illustrated in \FIGs{fg:mlwc} (a) and (b) as examples,
we have recently proposed a chemical bond-based ML model \cite{amano2024chemical,amano2025transferability} which predicts the location of Wannier centers for each bond or lone pair, and can be applied to a plentiful of organic molecules.

However, this model encounters difficulty with conjugated $\pi$ (aromatic) systems, because the electrons here are delocalized in the whole system and cannot be adequately described by classical Lewis structures \cite{lewis1916atom}.
The Wannier centers are also located at peculiar positions as illustrated by two N-methylaniline molecules in \FIGs{fg:mlwc} (c) and (d) as an example: 
Firstly, it can be observed that the Wannier centers reflect a Kekul\'e structure for the benzene ring, but the positions of double/single bonds are indeterminate [compare the double-bond positions in (c) and (d)] and the centers for the double bonds are significantly shifted towards the ring center compared with the typical double bonds as in cyclopentene [\FIG{fg:mlwc} (a)].
Additionally, the Wannier center on the N atom is somewhat weird because its lone pair is also conjugated with the benzene ring, and the red N--C bond in the figure has partial double bond characters.
These make it impossible to assign an integral bond order \textit{a priori}, which is the very first step in the bond-based model.
Given the ubiquity of conjugated molecules, a solution to this problem is highly needed.

Another approach for the dipole moment is to compute the Born effective charge (BEC) tensor \cite{ghosez1998dynamical}, defined as 
\begin{equation}\label{eq:becdef}Z_{I,{ij}}=\frac{\partial M_j}{\partial R_{Ii}}=\frac{\partial F_{Ii}}{\partial \mathcal E_{j}},\end{equation}
where $I$ is the index of atom, $i,j\in \{x,y,z\}$ and $F$ and $\mathcal E$ are interatomic force and external electric field, respectively.  
The change on dipole moment can then be readily computed by the integration of BEC.
Many ML models have been recently proposed to predict the BEC for each atom from the input structure \cite{kutana2025representing,schmiedmayer2024derivative}.
Since computation of BEC does not rely on the Lewis structures, conjugated systems can be treated as usual without any modification.
This approach shows good accuracy at IR range but is undertested in THz region.
In fact, the numerical noise/error can accumulate during the integration along the long MD trajectory and the accuracy of BEC models at low frequency might be questionable.
Moreover, it is computationally heavy to obtain BEC from DFT because one has to perform calculations at finite electric field along all three axes for finite difference using \EQ{eq:becdef}, and set very tight electronic convergence criterion to ensure numerical accuracy.

In this paper, we propose an atom-based dipole model that is conceptually similar to our previous bond-based model but works with the conjugated systems.
Here, an effective atomic dipole (EAD) for each atom is assigned by unitary transformation of the DFT wavefunction, and then predicted by a graph neural network (GNN).
It is demonstrated that the EAD model, while costing only $\sim$$1/3$ computational resources compared with the BEC models, shows similar or superior accuracy for dielectric function at THz or lower-frequency region.
The next section describes the computational method for EAD. In section \ref{sec:comptdetail}, we have applied our method to $\sim$500 types of organic molecules, followed by a discussion on properties of EAD and accuracy of ML models in section \ref{sec:discussion}. 

\section{Effective atomic dipole model}

\subsection{Model calculation}\label{sec:modelcalc}

\begin{figure*}
\includegraphics[scale=.6]{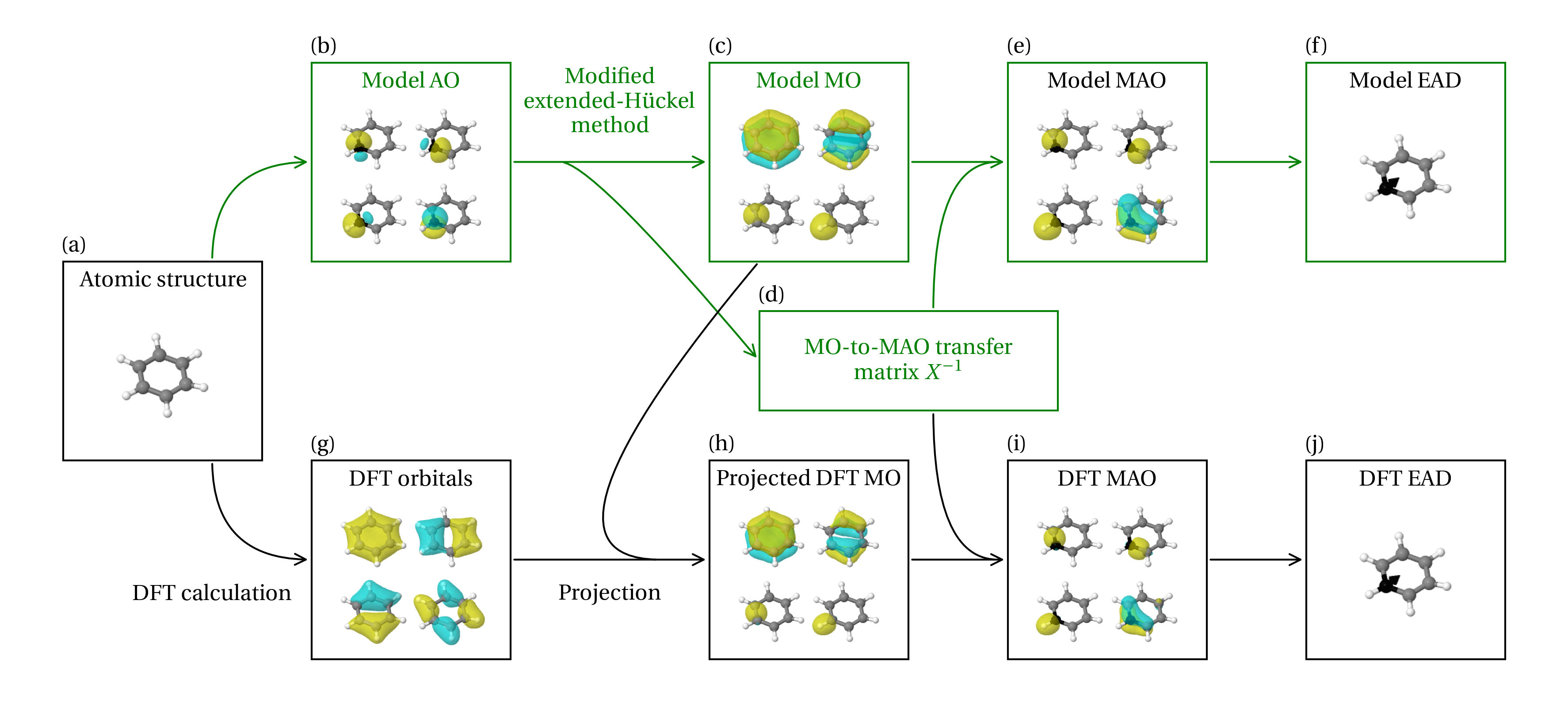}
\caption{Computation flow of effective atomic dipole (EAD).
(a) Input atomic structure, with a benzene molecule as an example.
(b) Construction of model atomic basis orbitals (AO). The iso-surfaces show the 4 basis functions of the black C atom, with yellow/light blue indicating $\pm$ sign.
(c) Computation of molecular orbitals (MO) from model calculation. The benzene molecule contains 1 conjugated $\pi$ system, 6 C--C and 6 C--H bonds in total. The upper panels show the two of the three occupied orbitals of the $\pi$ system with the lowest energy, and the lower panels show the bonding orbital of a C--C and a C--H $\sigma$ bond, respectively.
(d) Computation of the transfer matrix from MO to ``molecular-atomic orbitals'' (MAO, see text).
(e) MAO from the model calculation. The iso-surfaces show the 4 MAO of the black C atom, which have similar shape as the AO in (b).
(f) Computation of EAD from MAO. The direction and magnitude of EAD of the black carbon atom is represented by an arrow originating from the atom.
(g) DFT calculation. The 4 lowest Kohn--Sham orbitals are shown.
(h) Projection of DFT orbitals to obtain localized DFT orbitals with similar shape of model MO as in (c).
(i)(j) Computation of DFT MAO and EAD similar to (e) and (f), respectively.
}\label{fg:flow}
\end{figure*}

\FIG{fg:flow} shows the computation flow of the EAD. 
As the Kohn--Sham (KS) wavefunctions from DFT are spread Bloch functions, we need a Wannierization process to localize the wavefunctions to compute the EAD which is a local property.
For this purpose, we first perform a model calculation (upper line in \FIG{fg:flow} in green) to obtain a set of localized molecular orbitals (MO) consistent with the real electronic structure of the molecule. 
This model calculation also yields a semi-quantitatively correct model EAD.
Finally, we employ a one-shot projection scheme \cite{marzari2012maximally} to obtain the localized DFT wavefunction to compute the EAD at DFT level (lower line in black).

We work in the spirit of the H\"uckel approximation \cite{Huckel1931} for the model calculation, 
regarding all $\sigma$ bonds still as localized on the two bonding atoms,
but when multiple atoms connected by $\sigma$ bonds have $\pi$-electrons or lone pairs, all these electrons are treated as delocalized in the whole conjugated system.
When the conjugated system contains only 2 or 1 atoms, it automatically reduces to the usual double/triple bonds or lone pairs, respectively. 

We use Gaussian atomic orbitals (AO) as basis function for the model calculation, and construct one $s$ orbital and 4 orthonormal $sp^n$-hybrid orbitals at each H and $p$-block atom, respectively.
The shape of each $sp^n$ orbitals are determined by its bonding neighbors: One orbital pointing to each of the neighbors is assigned for forming $\sigma$ bonds, and the remaining orbitals are used for $\pi$ bonds or (possibly conjugated) lone pairs. \FIG{fg:flow} (b) shows the AOs for the black carbon atom in a benzene molecule as an example and the details about the basis sets are given in the Appendix \ref{seca:orb}.

In the next step, we calculate the model MO as the eigenfunction of the extended-H\"uckel Hamiltonian \cite{wolfsberg1952spectra}.
Here, the atomic valence-electron energies are assigned to the diagonal entries of the Hamiltonian, and the off-diagonal part is taken to be $H_{ij}=\frac{\textsub C{WH}}2 (H_{ii}+H_{jj}) \langle\chi_i|\chi_j\rangle$, where
$\textsub C{WH}$ is the Wolfsberg--Helmholz constant which we use a value of 1.5, and $\chi$ are the AO basis functions.
However, the eigenfunction of this Hamiltonian is not localized due to the coupling between different parts of the whole molecular system.
In order to obtain localized eigenfunction, we split the whole molecular system into subsystems of $\sigma$ bonds or conjugated $\pi$ systems according to the H\"uckel approximation, and set $H_{ij}=0$ between orbitals belonging to different subsystems to decouple them.
This effectively turns off the interaction between differnent subsystems, 
but in the meantime allows the eigenfunctions to be well localized within each subsystem [\FIG{fg:flow} (c)].
Only occupied MOs ($\textsub Qc$ orbitals with the lowest energy for a subsystem containing $2\textsub Qc$ electrons) are needed to compute EAD.
For each $\sigma$ bond, this is simply the the bonding orbital [lower panels in \FIG{fg:flow} (c) as two examples].
The two of lowest three $\pi$ orbitals for the benzene ring are shown in upper panels in \FIG{fg:flow} (c) as an example for conjugated systems.

\subsection{``Molecular-atomic orbitals''}\label{sec:mao}
While the MOs obtained in the last subsection are localized similar to Wannier functions,
they are still shared between several atoms.
To assign an EAD to each atom, we use linear combination of MOs to construct AO-like orbitals, referred to as ``molecular-atomic orbitals'' (MAO).
Similar to computing MO, construction of MAO is also performed for each subsystem individually.

Consider a subsystem formed by $N$ atoms and $2\textsub Qc$ electrons, with atom $I$ providing $2q_I$ electrons into the whole subsystem (the factor 2 comes from spin and is introduced for simpler notation below).
Since the MOs $\Psi_j$ are linear combination of the basis orbitals $\chi_{(Ii)}$ ($I$ and $i$ are index of atom and AO of the atom, respectively), i.e. $\Psi_j=\sum_{Ii} \chi_{(Ii)}X_{(Ii)j}$ with matrix $X$ formed by the lowest $\textsub Qc$ eigen vectors of the extended-H\"uckel Hamiltonian, it makes sense to compute the MAOs $\tilde \chi$ \emph{formally} by $\tilde \chi_{(Ii)}= \sum_j\Psi_j X^{-1}_{j(Ii)}$ for some matrix $X^{-1}$.
We note that $X^{-1}$ must satisfy the following conditions in order to define meaningful EADs:
\begin{enumerate}[nosep]
\item $\sum_{ji} |X^{-1}_{j(Ii)}|^2=q_I$, i.e., the atomic charges computed from MAO ($=2q_I-2\sum_{i} \langle\tilde \chi_{(Ii)}|\tilde \chi_{(Ii)}\rangle$) are neutral so that EAD can be defined,
\item $\sum_{Ii} X^{-1}_{j(Ii)}X^{-1}_{k(Ii)}=\delta_{jk}$, i.e., $X^{-1}$ is an isometry so that all physical quantities computed from MAO are the same from those computed from the MO $\Psi$.
\end{enumerate}
Here, we introduce positive undetermined scaling parameters $\lambda_I$ for each atom, and compute $X^{-1}$ similar to Moore–Penrose inverse as $X^{-1}:=VU^\top$ with $U$ and $V$ from the singular value decomposition (SVD) $\lambda_I X_{(Ii)j}=USV^\top$.
Because $U$ and $V$ are semi-orthogonal, the second condition above will be automatically satisfied, and the $\lambda_j$ parameters are then determined by the first condition above. 

In the special case of isolated lone pairs (not a part of any larger conjugated systems), the $X^{-1}$ is simply unity;
for $\sigma$ bonds (or general 2-center-2-electron bonds, $N/2=1=q_0=q_1$), $X^{-1}$ can also be explicitly solved to be $\frac1{\sqrt2}[1,1]$.
In more general cases, $\lambda_j$ and $X^{-1}$ can be computed numerically by the iterative method described in Appendix \ref{seca:mat} which yields a unique result.
The MAO computed in this way is found to have similar shape as AO, as shown \FIG{fg:flow} (e).

With MAO, we are ready to define the EAD for each atom at the level of model calculation (not DFT calculation) by  
\begin{equation}\label{eq:ead_def}\bm \mu_I:=2\sum_{i} \langle \tilde\chi_{(Ii)}|\bm r|\tilde\chi_{(Ii)}\rangle -Z_I\bm R_I,\end{equation}
where $\bm r$, $Z_I$ and $\bm R_I$ are position operator, valence charge and position of atom $I$, respectively, as shown in \FIG{fg:flow} (f).
Here, the MAO and the atom $I$ are refolded to the same minimum-distance periodic image, and
we use the convention that the dipole points from positive charges to negative charges.
For non-conjugated atom (have only $\sigma$ bonds or isolated lone pairs), the first term in \EQ{eq:ead_def} is simply $\sum_{j}^{\sigma \text{ bonds}} \langle\Psi_j|\bm r|\Psi_j\rangle + 2\sum_{j}^{\text{lone pairs}} \langle\Psi_j|\bm r|\Psi_j\rangle$, i.e., as if the electrons of the atom reside on the bonding orbital or lone pairs, which is intuitively correct.
It is important to note that our $\bm \mu_I$ is not a well-defined physical observable like BEC, because it depends on the parameters of the model calculation which carries arbitrariness.
We refer to this property as ``gauge dependence''. 
However, the sum of EADs of all atoms is well defined and equal to the correct total dipole moment due to our $X^{-1}$ being an isometry, i.e., 
\[\sum_I \bm \mu_I = 2\sum_j \langle \Psi_j|\bm r|\Psi_j\rangle -\sum_I Z_I\bm R_I+\bm P,\]
with $\bm P$ certain integer multiple of the polarization quanta that comes from the refolding of the MAOs.

\subsection{From model calculation to DFT calculation}\label{sec:svdproj}
The MAO and EAD from the model calculations above are only semi-quantitatively correct. 
However, to obtain the EAD at DFT level, it suffices to replace the localized MOs $\Psi$ above with those computed from DFT (but the matrix $X^{-1}$ from model calculation are used as is).
This is accomplished by a one-shot projection scheme \cite{marzari2012maximally}.
Let $\Psi^{\text{Model}}$ and $\Psi^{\text{DFT}}$ be the model MOs and (possibly not localized) occupied DFT wavefunctions, respectively, and compute the SVD of overlap matrix $M_{ij}=\langle \Psi_i^{\text{DFT}}|\Psi_j^{\text{Model}}\rangle=USV^\top$. 
Then, the set of orbitals $\sum_i (UV^\top)_{ij}\Psi_i^{\text{DFT}}$ are orthonormal, span the same space as $\Psi_i^{\text{DFT}}$, and have similar shape as the model orbitals [see \FIG{fg:flow} (h)]. Therefore, these orbitals can be used to compute the DFT EADs, which also sum up to the correct dipole moment from DFT (up to an integer multiple of polarization quanta).


\subsection{Computational cost}\label{sec:cost}
In subsection \ref{sec:modelcalc}, we have decomposed the whole molecular system using H\"uckel approximation.
This not only make the eigen function of the model Hamiltonian localized, but also dramatically speedup the computation.
This is because instead of diagonalizing one big $O(\textsub N{atom}\times\textsub N{atom})$ Hamiltonian matrix where $\textsub N{atom}$ denotes the total number of atoms, we only need to diagonalizing $O(\textsub N{atom})$ number of $O(1\times1)$ matrices, reducing the computational cost from $O(\textsub N{atom}^3)$ to $O(\textsub N{atom})$.
The same cost reduction also applies to the computation of $X^{-1}$ matrices.

In fact, we have found that running all the model calculation [the green part in \FIG{fg:flow}] is actually about an order of magnitude faster than executing the neural network for predicting EADs (subsection \ref{sec:nn}).
The only computationally heavy part is the projection of DFT wavefunctions [subsection \ref{sec:svdproj} and \FIG{fg:flow} (h)] which costs $O(\textsub N{atom}^3)$. However, this is still much faster than the DFT calculation itself and actually not required if one only calculate the model EADs.

\section{Computation details} \label{sec:comptdetail}
\subsection{Machine learning models}\label{sec:nn}

We have designed a GNN model inspired by CHGNet model \cite{deng2023chgnet} and a covariant model \cite{zhang2020deep} to predict the EAD from atomic species and positions.
The total dipole moment is then just the sum of EADs of all atoms.
This model contains $\sim$315k trainable parameters and automatically ensures the required translational, rotational and permutational invariance (See Appendix \ref{seca:nn} for detail of the model).
While a relatively simple neural network was employed in our previous work \cite{amano2024chemical}, it is found that for conjugated molecules, using GNN is strictly required to reach high accuracy (see subsection \ref{sec:ead})

The BEC can be also directly predicted using a GNN with similar architecture \cite{kutana2025representing}.
However, the real BEC should be a total differentiation, which is not enforced by such GNN.
This differential error is usually harmless for dielectric properties at IR region \cite{kutana2025representing,schmiedmayer2024derivative}.
However, at THz range, such numerical error accumulates over time, and the total dipole moment from integrating BEC will drift away from the correct value.
We have found that this alone (without error from machine learning models) is already catastrophic as calculating the dielectric function even with DFT BEC is not feasible (see subsection \ref{sec:diel}).
However, we have also found that the solution to this problem is to use the strategy proposed in Ref. \cite{schmiedmayer2024derivative} of expressing the BEC as the derivative of the total dipole moment predicted by the GNN, and use the dipole moment to evaluate dielectric function.
This also allows us to use exactly the same GNN architecture for predicting BEC, which is beneficial for comparison of performance. 

We have trained two types of ML model using the following unified $L1$ loss
\begin{equation}\label{eq:loss}L=a\left|\sum_I(\bm \mu_I^\text{true}-\bm \mu_I^\text{pred})\right|+
b\sum_I|\bm \mu_I^\text{true}-\bm\mu_I^\text{pred}|+
c\sum_I\left|\bm Z_I^\text{true}-\nabla_{\bm R_I}\sum_J\bm \mu_J^\text{pred}\right|_{\text{Frobenius}},\end{equation}
where $\bm \mu_I$ and $\bm Z_I$ denote the EAD and BEC of atom $I$, respectively; the superscript ``true'' and ``pred'' denotes DFT values and those predicted by ML models; and finally $a$, $b$ and $c$ are all hyper-parameters.
In the first model, referred to as EADNN model, we set $a=b=1$ and $c=0$ (the unit is \si{\angstrom} and \si{e\angstrom} for distance and dipole moment, respectively), i.e. train on EAD and total dipole moment,
while for the second type, referred to as the BECNN model, is similar to that proposed in Ref. \cite{schmiedmayer2024derivative} with $a=c=1$ and $b=0$, i.e. train on the BEC instead of EAD.
It is found that including the total dipole moment ($a=1$) in the loss reduces the error on the total dipole moment \emph{on the validation set} slightly for EADNN (by about 1/10) and significantly (by about 2/3) for BECNN compared with $a=0$.
Both models are written with TensorFlow \cite{tensorflow} and trained using the Adamax optimizer \cite{kingma2014adam} with weight decay.

The costs for running both EADNN and BECNN scale linearly with total number of atoms.
However, due to requirement to compute the derivative with respect to the atomic coordinates, running BECNN is found to be about 3 times slower compared with the EADNN.

\subsection{Construction of ML models}\label{sec:dft}

The ML models are trained and tested on a database containing about 500 types of organic molecules, mainly from the Melendey solvent database \cite{mlist}.
For each molecule, we have generated $\sim$1000 independent atomic structures in liquid phase containing about 360$\sim$400 atoms by a \SI{50}{ps} classical MD simulations starting from randomized molecular configurations.
The MD simulation is performed using the LAMMPS \cite{thompson2022lammps} and GROMACS \cite{abraham2015gromacs} packages with the GAFF2/AM1BCC force field \cite{he2020fast,jakalian2002fast} and Langevin NVT thermostat at \SI{300}{K} and the experimental density.

DFT calculations are then performed with the CPMD package \cite{cpmd} for all the structures.
We have used GTH norm-conserving pseudopotentials \cite{goedecker1996separable} with the BLYP exchange-correlation functional \cite{lee1988development,becke1988density},
and expanded the electronic wavefunctions using planewaves with a cutoff of \SI{100}{Ry}.
The output wavefunctions are finally passed to a homemade code which calculates the EAD using the algorithm described above.

To compare the performance between EADNN and BECNN, We have also computed the BEC for a subset of the structures ($\sim$550 structures each for all 70 hydrocarbons in the database and $\sim$900 structures for each of 33 selected molecules containing also N and O), by additional 6 DFT calculations at a finite electric field of \SI{\pm3e-4}{au} along each axis. 
Here, the energy convergence condition has been set to a very tight value of \SI{e-8}{au} for the accuracy of BEC (the default \SI{e-5}{au} is already sufficient for EAD). 

\subsection{Dielectric functions}\label{sec:dfcalc}

The complex dielectric function of liquid can be computed by the autocorrelation function of dipole moment \cite{kubo1957statistical,neumann1983calculation} as
\begin{equation}\label{eq:dfunc_dip}\varepsilon(\omega)=\frac{\langle \bar{\bm M}^2\rangle}{3\textsub kBTV}+\varepsilon(\infty)-\frac{\ii\omega}{3\textsub kBTV}\int_0^\infty K_{\bar{\bm M}\bar{\bm M}}(t)\exp(-\ii\omega t)\,\dd t,\end{equation}
where $\textsub kB$, $T$, $V$ and $\varepsilon(\infty)$ denote the Boltzmann constant, temperature, volume, and high frequency dielectric function, respectively; $\bm M=\sum_I\bm \mu_I$ is the dipole moment of the whole simulation cell; $\bar{\bm M}(t)=\bm M(t)-\langle \bm M\rangle$ is the zero-mean dipole moment, and $K_{\bm X\bm Y}(t)=\langle\bm X(t)\cdot\bm Y(0) \rangle$ denotes time correlation function between vector variables $\bm X$ and $\bm Y$.

Instead of using dipole moment, one can also use the autocorrelation function of time derivative of dipole moment, which is directly related to the BEC by $\dot {\bm M}=\sum_I\bm Z_I\cdot \bm v_I$, with $\bm v_I$ the velocity of atom $I$, and 
\begin{equation}\label{eq:dfunc_bec}\varepsilon(\omega)=\frac{\langle \bar{\bm M}^2\rangle}{3\textsub kBTV}+\varepsilon(\infty)-\frac{\ii}{3\textsub kBTV\omega}\int_0^\infty K_{\dot{\bm M}\dot{\bm M}}(t)\exp(-\ii\omega t)\,\dd t.\end{equation}

We use the snap shots sampled on regular intervals from MD trajectories for computing the dielectric function.
As the highest frequency component in the trajectory is the O--H stretching at $\sim$\SI{100}{THz}, the largest allowed sampling interval is about $\SI 4{fs}$.
The length of the trajectory required to capture the low-frequency dielectric properties for polar molecules is far beyond the possibility of DFT calculation.
In subsection \ref{sec:diel} below, we use a maximum of \SI{200}{ps} trajectory, which is sufficient for evaluation of the ML models against DFT calculations. 

\section{Results and discussions}\label{sec:discussion}
\subsection{Properties of EAD}\label{sec:ead}

\begin{figure}
\includegraphics[scale=.6]{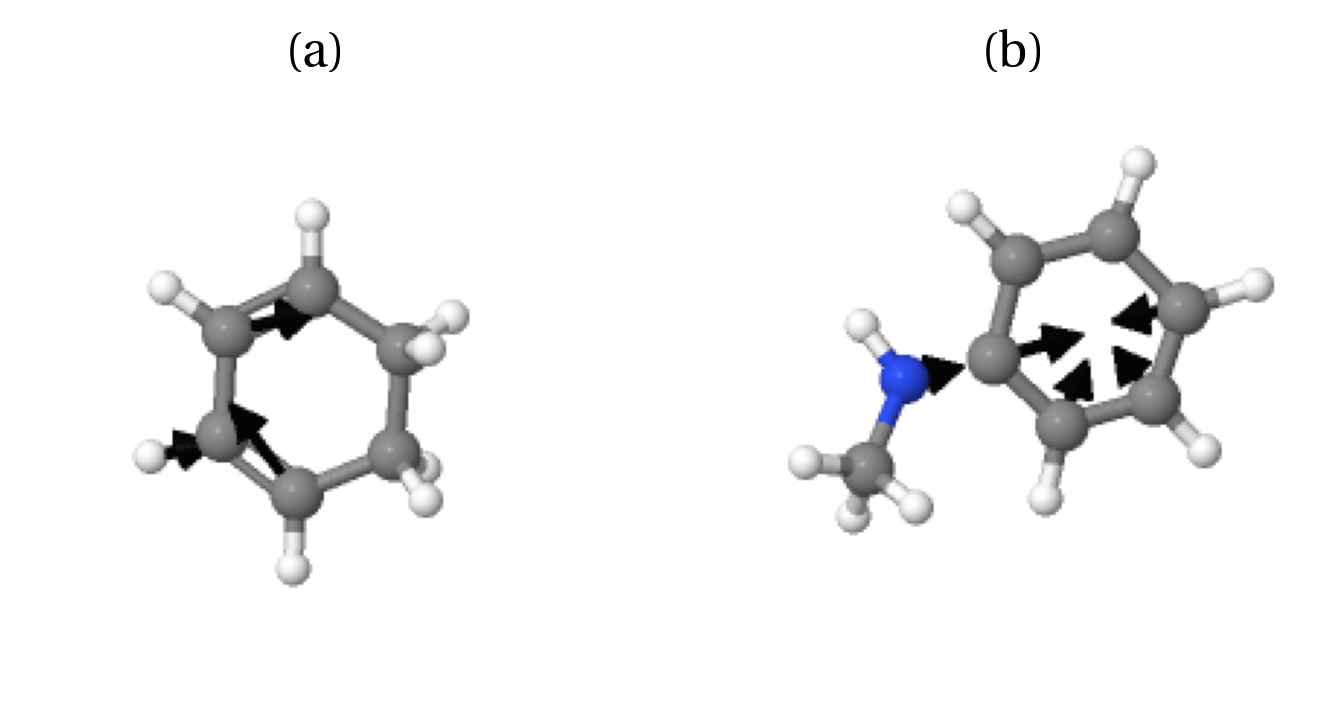}
\caption{(a) The EAD of some atoms in a 1,3-cyclohexadiene molecule. The direction and magnitude of each EAD is represented by an arrow originating from the atom. (b) The same as (a) but in a N-methylaniline molecule.}\label{fg:ead}
\end{figure}

\begin{figure}
\includegraphics[scale=.6]{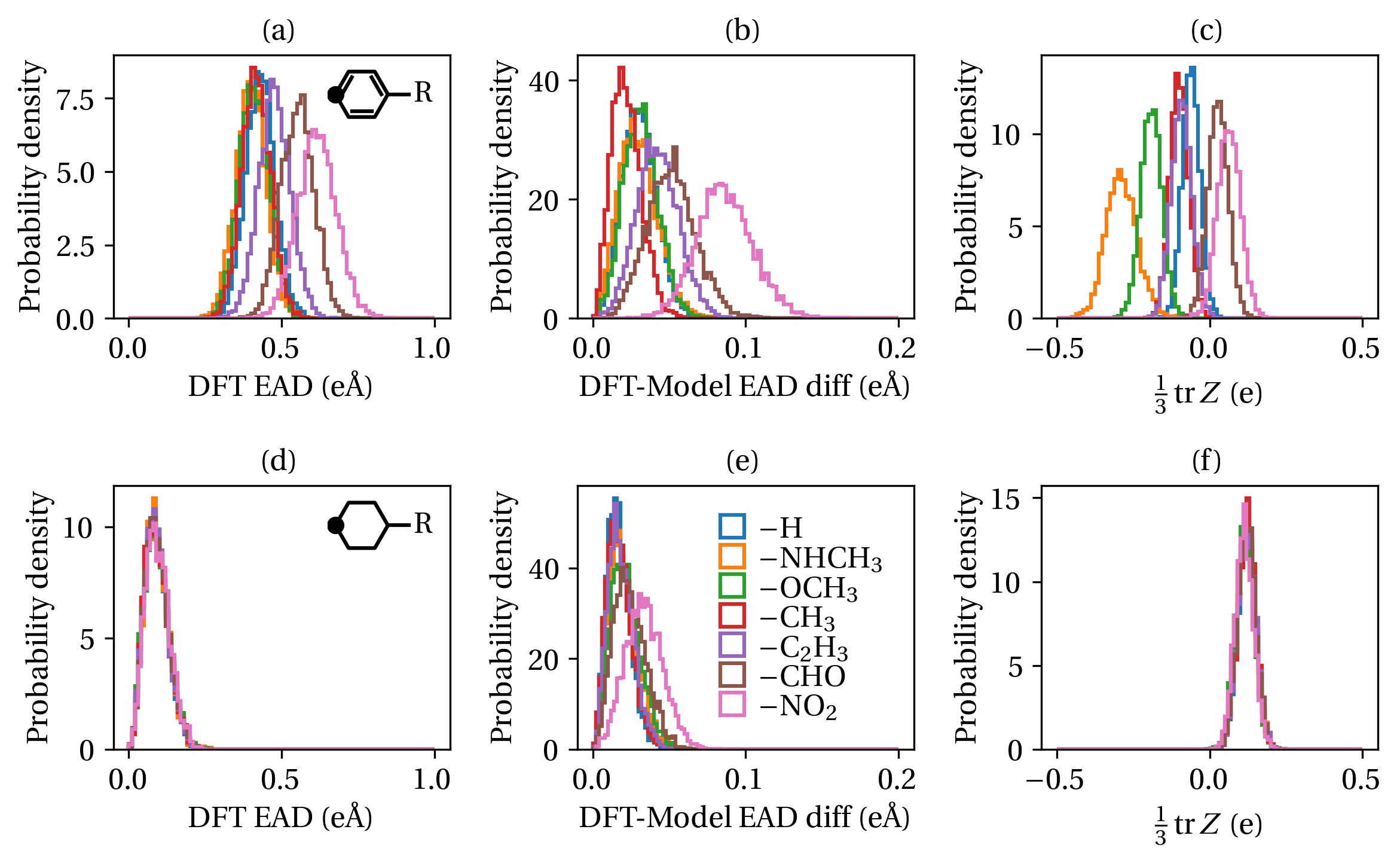}
\caption{(a) The distribution of norm of the DFT EAD of the C atoms at para position in several singly substituted benzene (the atom marked by the dot in the inset).
(b) The distribution of norm of difference between DFT and model EADs of the same atoms.
(c) The distribution of the trace of DFT BEC of the same atoms.
(d)(e)(f) The same as (a), (b) and (c), respectively, but the phenyl groups in the molecules are replaced by cyclohexyl groups.
}\label{fg:dstr}
\end{figure}

\begin{figure}
\includegraphics[scale=.6]{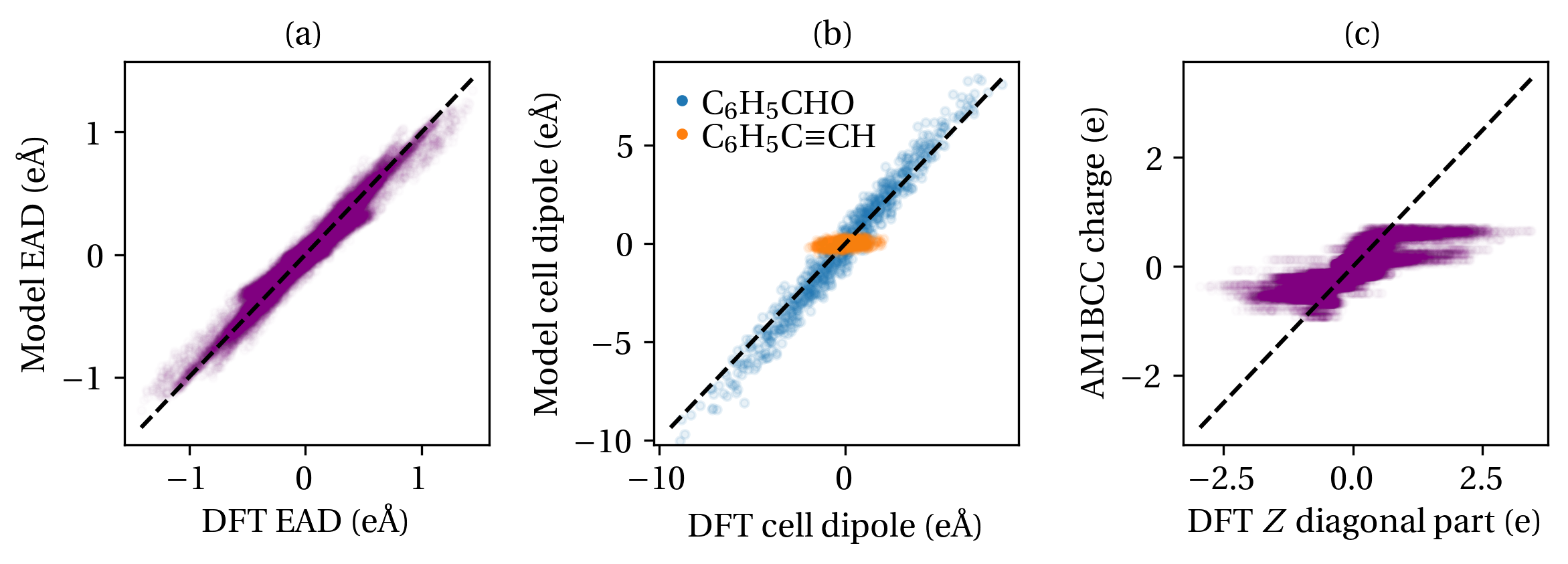}
\caption{(a) Parity plot between the EAD computed from DFT and model calculation.
(b) Parity plot between the total cell dipole computed from DFT and model calculation of benzaldehyde and phenylacetylene in liquid phase containing 25 molecules, respectively.
(c) Parity plot between diagonal part of the Born effective charge $Z$ and AM1BCC charge.
}\label{fg:comp}
\end{figure}

We have qualitatively explored the orientation of EAD and found that it generally points to the electron-rich part near the atom, 
which is the bond center for terminal atoms such as H in \FIG{fg:ead} (a) (and also halogens and O atoms in carbonyl groups), and double bonds (or $\pi$ electrons) for non-terminal atoms (the EAD of saturated non-terminal atoms is found to be very small). 
The conjugate effect also strongly affects the direction of EAD.
In 1,3-cyclohexadiene [\FIG{fg:ead} (a)], the C=C--C=C structure is only partially conjugated, and the EADs of the 4 C atoms still approximately point towards the center of the double bonds.
However, in completely conjugated structures as benzene ring in N-methylaniline molecule, as all C--C bonds are mostly equivalent, the EAD of C atoms points towards the ring center as shown in [\FIG{fg:ead} (b)].
The EAD of N atom here also points towards the benzene ring, showing effect of the $p$--$\pi$ conjugation.

It is also noticed that the EAD in conjugated system has very large substituent effect in a way similar to the aromatic electrophilic substitution reaction.
\FIG{fg:dstr} (a) shows the distribution of the norm of EAD of C atom at para position (the one marked by black dot in the inset of figure) of several singly substituted benzene derivatives.
Despite having identical chemical environment up to at least 3\textsuperscript{rd} neighbors, the EAD of these atoms differ significantly and
are generally larger when the substituent is an electron withdrawing group (EWG, such as --NO\ts2) as it attracts the electron from the conjugated system.
We have found that the EWG can also make the BEC of these atoms much more positive as shown in \FIG{fg:dstr} (c), and therefore such dependence is not an artifact of ``gauge dependence'' of EAD (since BEC is gauge independent).
On the contrary, when the benzene ring is replaced by cyclohexyl group, such long-range dependence all disappears [\FIGs{fg:dstr} (d) and (f)].
This indicates that the properties of conjugated system are very sensitive to the surrounding environment due to the delocalization effect and far more difficult to predict with high accuracy.

\FIG{fg:comp} (a) shows the parity plot between the EAD from the model calculation and DFT for all structures in the database (the subset that BEC are available as described in subsection \ref{sec:dft}).
It is revealed that the model EAD for individual atoms can actually serve as a good predictor for DFT EAD.
As illustrated in \FIG{fg:comp} (b), the total cell dipole computed from model EAD is also semi-quantitatively correct for polar molecules such as benzaldehyde,
but completely fails for non-polar molecules such as phenylacetylene.
Additionally, the model EAD can only partially predict the substituent effect on benzene ring discussed above because the difference between DFT and model EADs still shows non-negligible dependence on the substituent as shown in \FIG{fg:dstr} (b).
These failures are evidently due to omission of the many-body induction effects in the model calculation as all interactions between subsystems are turned off to obtain localized model MOs (subsection \ref{sec:modelcalc}), and indicates the absolute necessity of using delicate ML models such as GNN to accurately predict these quantities.

However, given the good agreement between DFT and model EADs and the fact that computation of model EAD is very fast (subsection \ref{sec:cost}) suggests that instead of training on the bare DFT EAD, it is much more efficient to perform delta training, i.e., training on the difference between in DFT and model EADs. We have found that this is indeed the case and an error reduction of 30\% can be achieved.
However, in training of the BECNN, we did not perform the delta training because of the large discrepancy between semi-empirical charge and BEC as shown in \FIG{fg:comp} (c) and the fact that the off diagonal component of BEC is also rather large (i.e., strong anisotropy) in the molecular systems. More importantly, assigning non-integral charge is not fully compatible with the quantized dipole moment in periodic cells.

\subsection{Validation of ML models}
\begin{figure}
\includegraphics[scale=.6]{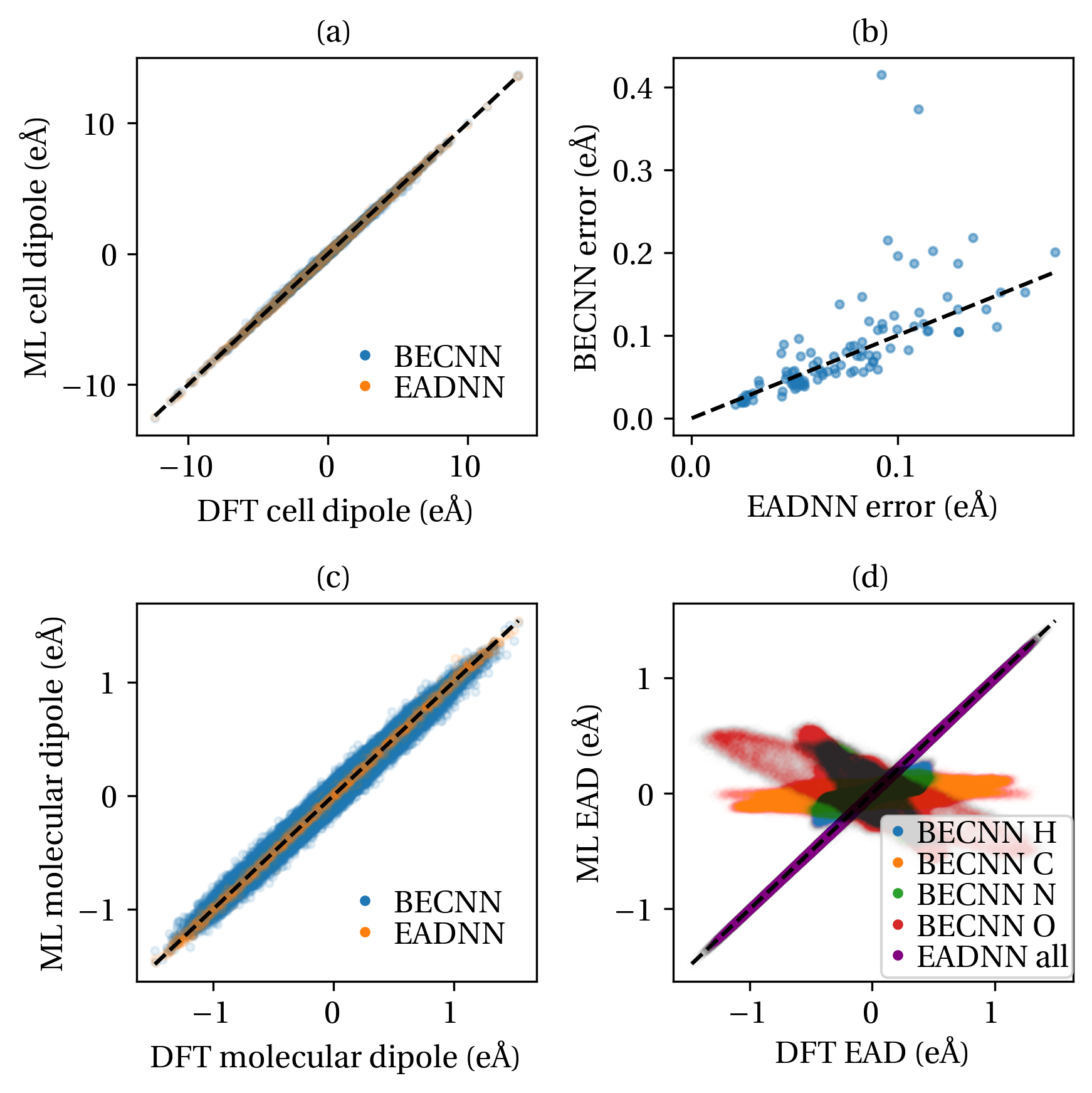}
\caption{Validation result of ML models. (a) The parity plot between DFT cell dipole ($x$, $y$ or $z$ component) and the predicted values by BECNN and EADNN, on the validation data set.
(b) Comparison between BECNN and EADNN of the mean absolute error in cell dipole for different molecules.
(c) Comparison between the molecular dipole moment computed from DFT Wannier function and the prediction by BECNN and EADNN.
(d) The same as (c) but for EAD.
}\label{fg:train}
\end{figure}

While the EAD is important for training the ML model, it is not a valid physical observable (not gauge independent) and we need to use the total cell dipole, the only physically meaningful quantity, to evaluate the models.
\FIG{fg:train} (a) shows the validation parity plot of total cell dipole moment of the EADNN and BECNN trained using the same dataset (the subset with BEC).
We find that both models show excellent accuracy of 0.09 and \SI{0.11}{e\angstrom} in mean absolute error (MAE) for EADNN and BECNN, respectively, on the cell dipole moment.
Comparatively, the BECNN is not as consistent as EADNN, giving large errors on some particular molecules as shown in \FIG{fg:train} (b).
We have also found that the MAE of cell dipole of EADNN on the whole dataset is \SI{0.11}{e\angstrom} albeit containing almost 5$\times$ molecule types, indicating a good generalizability of the GNN model.

It is interesting that the two models decompose the cell dipole quite differently.
The EADNN is guided by the loss to predict the EADs and inexplicitly the molecular dipole moments to the DFT values, and shows very high accuracy also on both quantities.
However, this is not enforced in the training of BECNN,
which is found to ``define'' molecular dipole moments slightly differently as shown in \FIG{fg:train} (c).
Moreover, the predicted EADs are completely different between both models [\FIG{fg:train} (d)], and even the directions are opposite for O atoms.
This is because the easiest feature for the ML model to capture near O atoms is presumably the overall tendency of the dipole moment to point toward O atoms due to their high electronegativity. However, our EAD model treats O atoms as sharing electrons through covalent bonds, and the EAD points from the atom toward the bond centers, which is opposite to the direction of the overall dipole moment.

\subsection{Dielectric function}\label{sec:diel}

\begin{figure}
\includegraphics[scale=.6]{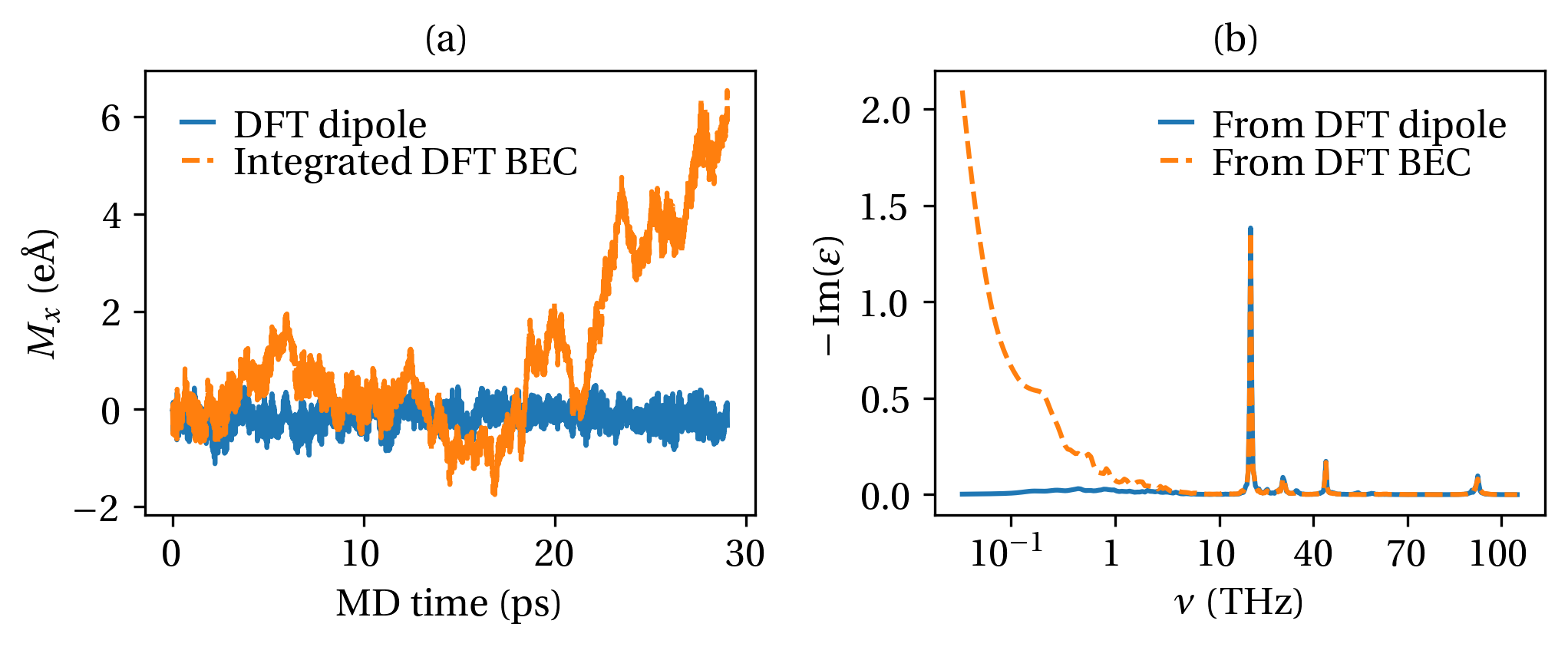}
\caption{(a) The $x$ component of total cell dipole moment of liquid benzene along a MD trajectory of 30 molecules. The DFT dipole moment is computed using the Berry phase.
(b) Imaginary part of the dielectric function computed from autocorrelation of dipole moment and time-derivative of dipole moment via Born effective charge. 
The linear frequency $\nu=\frac\omega{2\pi}$ and the horizontal axes are logarithmic and linear below and above \SI{10}{THz}, respectively.
}\label{fg:dplbec}
\end{figure}

As stated in subsection \ref{sec:dfcalc}, the dielectric function can be evaluated from either cell dipole or BEC.
While the two expressions are mathematically equivalent, with numerical noises, the calculation by cell dipole is much more robust at low frequency.
\FIG{fg:dplbec} (a) shows the cell dipole moment along a \SI{29}{ps} step MD trajectory of liquid benzene containing 30 molecules with sampling interval \SI{20}{au} (\SI{0.48}{fs}) computed from Berry phase and integrated using BEC (both from DFT calculations).
At long time scale, we have found that the numerical noises in BEC accumulate and the integrated BEC diverges similar to a random walk.
This translates to large errors in the low frequency part of the dielectric function calculated using BEC [\EQ{eq:dfunc_bec}] as shown in \FIG{fg:dplbec} (b), indicating that ML models which directly predict the BEC (without expressing BEC as the derivative of dipole moment) will fail at THz region.
In the BECNN, by virtue of derivative training, the total dipole moment is directly available and does not suffer from the divergence problem, making it usable for computing the dielectric function.

\begin{figure}
\includegraphics[scale=.6]{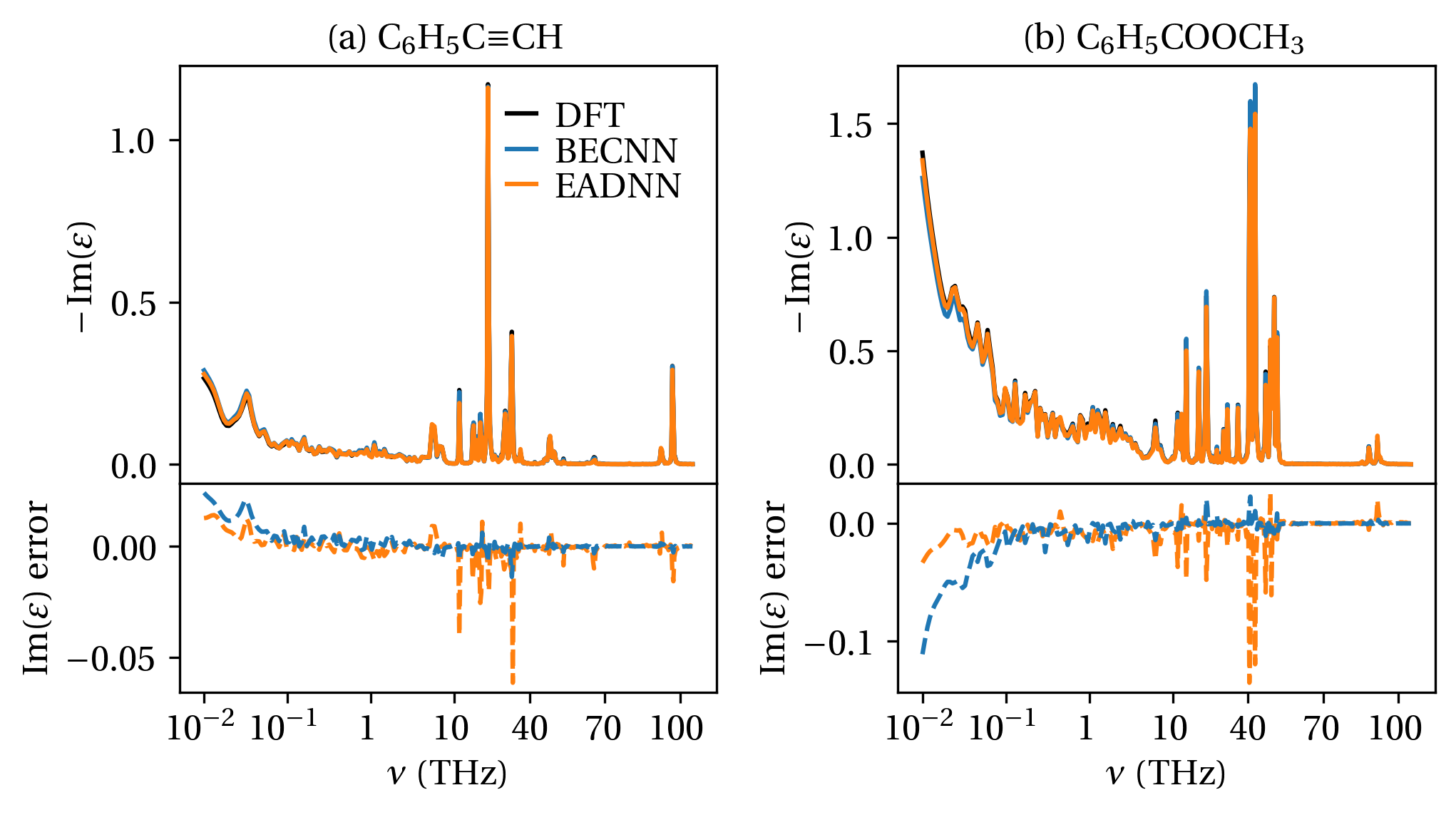}
\includegraphics[scale=.6]{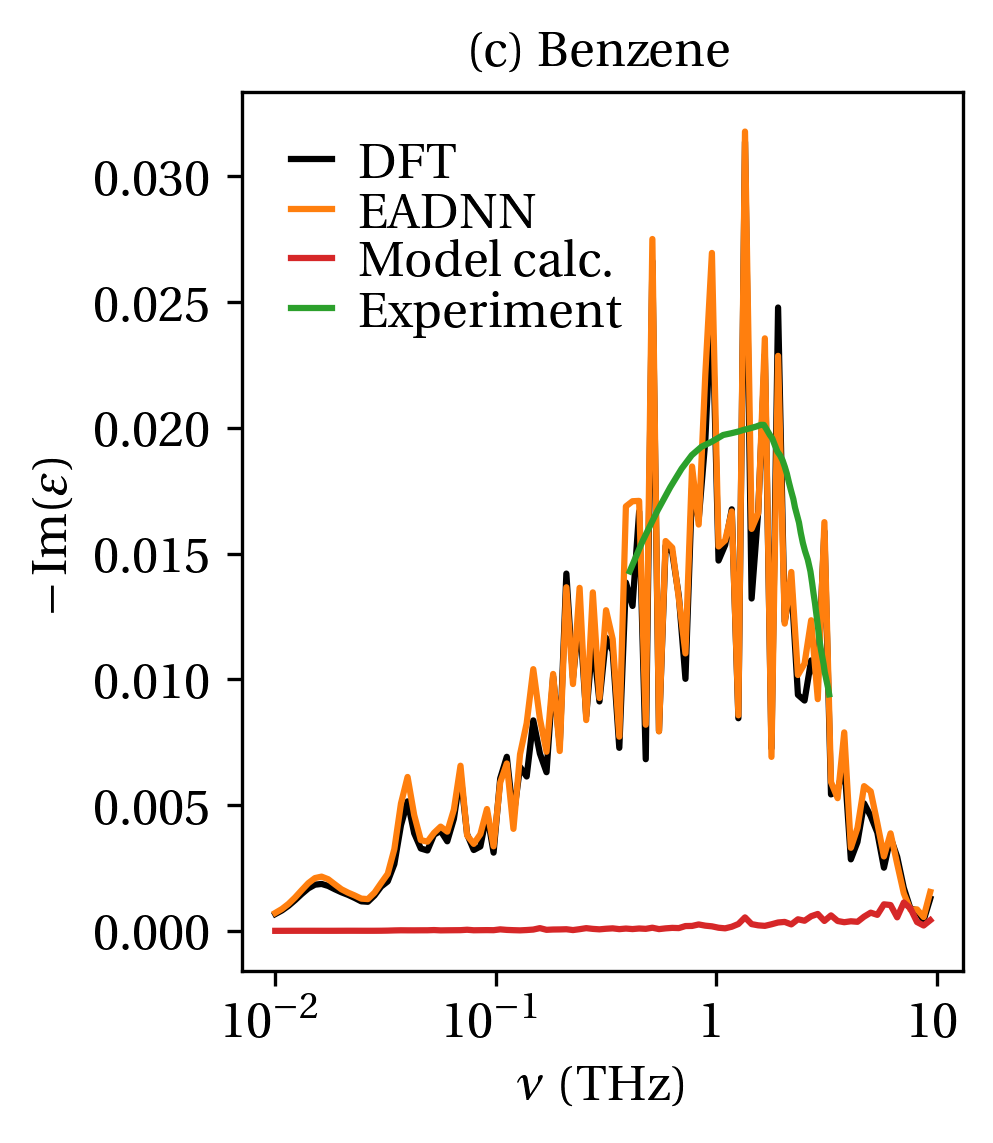}
\caption{Imaginary part of dielectric function computed using autocorrelation of DFT and ML-predicted dipole moment from a \SI{200}{ps} MD trajectory with \SI2{fs} sampling interval of liquid (a) phenylacetylene containing 25 molecules, (b) methyl benzoate containing 20 molecules and (c) benzene containing 30 molecules. The DFT dipole moments are computed from the Berry phase and the lower panels in (a) and (b) show the deviation of the BECNN and EADNN compared with DFT.
In (c), the results using dipole moments from the model calculation and experiment \cite{ro2000temperature} are also included.
}\label{fg:dfunc}
\end{figure}

However, we have then found that for predicting the dielectric constant at (sub-)THz range, BECNN is still not as accurate compared with EADNN which is directly trained on the dipole moment.
\FIGs{fg:dfunc} (a) and (b) shows the dielectric function of phenylacetylene (nearly non-polar) and methyl benzoate (polar), respectively, calculated using dipole moments along a \SI{200}{ps} classical MD trajectory with sampling interval \SI2{fs}.
It is important to note that the absolute value for dielectric functions calculated here is not reliable at for frequencies below $\sim$\SI{0.1}{THz} due to under sampling but our purpose here is to evaluate the two ML models using the DFT result.
It is clear that while both models show excellent accuracy compared with DFT results at the whole frequency region,
the EADNN generally reproduces the DFT result better than BECNN at (sub-)THz range.
On the other hand, at IR range, BECNN becomes superior than EADNN.

This difference is presumably because as BECNN is trained on the derivatives of the dipole moment, 
it can capture the change of dipole moment induced by small atomic displacement better than EADNN.
Such sensitivity is exactly the requirement to predict the dielectrics at IR range which is dominated by intramolecular vibrations.
However, at THz region, the dielectrics are mostly affected by collective molecular movements such as libration, which are generally accompanied by relatively large atomic displacements, causing the error on BECNN to accumulate (though not that severe to randomly drift and diverge).
On the other hand, the EADNN is trained to predict the dipole moments of a diversity of structures and therefore superior to predict the correlation between very different structures.

Comparing with experimental results, it is also demonstrated that many-body polarization effects are highly important at THz range as illustrated in \FIG{fg:dfunc} (c). 
The imaginary part of $\varepsilon(\omega)$ of benzene measured experimentally \cite{ro2000temperature} has a peak at $\nu=\frac{\omega}{2\pi}\approx\SI{2}{THz}$.
However, the EAD model calculation, which does not include any information from inter-molecular interactions, gives rise to a nearly vanishing $\Im(\varepsilon)$,
and the experimental result can only be quantitively reproduced by carefully treating this collective polarization effects as in DFT or ML models.

\subsection{Comparison and generalization}

We have summarized a comparison between EADNN and BENCC in \TBL{tb:comp}.
In practice, while both models can be employed to predict dielectric functions, 
considering that our EADNN model is about 3 times faster to train than traditional BECNN (for both for DFT calculation and NN training),
it will generally be a more flexible choice.
This is particularly true if the interest lies in the low-frequency region where EADNN is also more accurate.
It seems that a better model can be trained by adding the loss for EAD and BEC [$a=b=c=1$ in \EQ{eq:loss}], but we have found that this is not the case and such mixing results in large degradation of accuracy due to incompatibility of two models. 

\begin{table}
\caption{Comparison between the performance of BECNN and EADNN.}\label{tb:comp}
\begin{tabular}{ccc}
\hline
\hline
Model & EADNN & BECNN\\
\hline
Gauge independent? & No & Yes \\
Training cost & Low & $\sim$$3\times$\\
Dielectrics below THz range & Better &\\
Dielectrics at IR range & & Better\\
\hline
\hline
\end{tabular}
\end{table}

Given the similar relation that the BEC is the derivative of total dipole moment and that the force is the negative derivative of total energy,
we find that the training of BECNN is highly similar to that of ML interatomic potentials: The first and last terms in \EQ{eq:loss} correspond to the total energy and force, respectively.
This indicates that, conversely, a new type of ML potential may be trained in a similar fashion to EADNN.
In fact, by using the MAO, the DFT total energy as a functional of electronic wavefunctions can be systematically written as the sum of atomic self-energies and interaction energies between atoms,
and these local quantities can be directly learned by neural networks.
Considering the difference between EADNN and BECNN, we expect this kind of model to be more suitable than conventional ML potentials for comparing the energies between different very structures (e.g. between different phases).

\section{Conclusion}
In conclusion, we have constructed a method to systematically assign a dipole moment to each atom in molecular systems, and the sum of EAD is equal to the true dipole moment computed by DFT.
This opens up a possibility to train ML models for predicting dipole moment from atomic structures and also dielectric function using long MD trajectories,
and works even for conjugated molecules where previous bond-based model does not work.
EAD depends strongly on the local environment but can be accurately predicted by using GNN models.
For evaluation of the complex dielectric function $\varepsilon(\omega)$, it is found that compared with BEC-based ML training methods, our method only cost 1/3 of the computational resources but performs better at (sub-)THz region and only slightly worse at IR range, 
providing a competitive way to analyze and design dielectric materials.

\begin{acknowledgments}
This work used computational resources of Fugaku supercomputer provided by RIKEN HPCI System Research Project (Project ID: hp250046), and the facilities of the Supercomputer Center, the Institute for Solid State Physics, The University of Tokyo.
\end{acknowledgments}

\bibliography{rc.bib}

@article{Huckel1931,
  author    = {Erich H{\"u}ckel},
  title     = {Quantentheoretische Beitr{\"a}ge zum Benzolproblem. I. Die Elektronenkonfiguration des Benzols und verwandter Verbindungen},
  journal   = {Zeitschrift f{\"u}r Physik},
  year      = {1931},
  volume     = {70},
  number     = {3--4},
  pages      = {204--286},
  publisher  = {Springer},
}

@article{lewis1916atom,
  title={The atom and the molecule.},
  author={Lewis, Gilbert N},
  journal={Journal of the American Chemical Society},
  volume={38},
  number={4},
  pages={762--785},
  year={1916},
  publisher={ACS Publications}
}

@article{amano2024chemical,
  title={Chemical bond based machine learning model for dipole moment: Application to dielectric properties of liquid methanol and ethanol},
  author={Amano, Tomohito and Yamazaki, Tamio and Tsuneyuki, Shinji},
  journal={Physical Review B},
  volume={110},
  number={16},
  pages={165159},
  year={2024},
  publisher={APS}
}

@article{amano2025transferability,
  title={Transferability of the chemical-bond-based machine learning model for dipole moment: The GHz to THz dielectric properties of liquid propylene glycol and polypropylene glycol},
  author={Amano, Tomohito and Yamazaki, Tamio and Matsumura, Naoki and Yoshimoto, Yuta and Tsuneyuki, Shinji},
  journal={Physical Review B},
  volume={111},
  number={16},
  pages={165149},
  year={2025},
  publisher={APS}
}

@inproceedings{siegel2004terahertz,
  title={Terahertz technology in biology and medicine},
  author={Siegel, Peter H},
  booktitle={2004 IEEE MTT-S International Microwave Symposium Digest (IEEE Cat. No. 04CH37535)},
  volume={3},
  pages={1575--1578},
  year={2004},
  organization={IEEE}
}

@article{neumann1983calculation,
  title={On the calculation of the frequency-dependent dielectric constant in computer simulations},
  author={Neumann, M and Steinhauser, O},
  journal={Chemical physics letters},
  volume={102},
  number={6},
  pages={508--513},
  year={1983},
  publisher={Elsevier}
}

@article{deng2023chgnet,
  title={CHGNet as a pretrained universal neural network potential for charge-informed atomistic modelling},
  author={Deng, Bowen and Zhong, Peichen and Jun, KyuJung and Riebesell, Janosh and Han, Kevin and Bartel, Christopher J and Ceder, Gerbrand},
  journal={Nature Machine Intelligence},
  volume={5},
  number={9},
  pages={1031--1041},
  year={2023},
  publisher={Nature Publishing Group UK London}
}

@article{fu2025learning,
  title={Learning smooth and expressive interatomic potentials for physical property prediction},
  author={Fu, Xiang and Wood, Brandon M and Barroso-Luque, Luis and Levine, Daniel S and Gao, Meng and Dzamba, Misko and Zitnick, C Lawrence},
  journal={arXiv preprint arXiv:2502.12147},
  year={2025}
}

@article{bone2024new,
  title={A new method to calculate broadband dielectric spectra of solvents from molecular dynamics simulations demonstrated with polarizable force fields},
  author={Bone, Rebecca A and Chung, Moses KJ and Ponder, Jay W and Riccardi, Demian and Muzny, Chris and Sundararaman, Ravishankar and Schwarz, Kathleen},
  journal={The Journal of Chemical Physics},
  volume={161},
  number={6},
  year={2024},
  publisher={AIP Publishing}
}

@article{nymand2001temperature,
  title={The temperature dependent dielectric function of liquid benzene: Interpretation of THz spectroscopy data by molecular dynamics simulation},
  author={Nymand, Thomas M and R{\o}nne, Cecilie and Keiding, S{\o}ren R},
  journal={The Journal of Chemical Physics},
  volume={114},
  number={12},
  pages={5246--5255},
  year={2001},
  publisher={American Institute of Physics}
}

@article{marzari2012maximally,
  title={Maximally localized Wannier functions: Theory and applications},
  author={Marzari, Nicola and Mostofi, Arash A and Yates, Jonathan R and Souza, Ivo and Vanderbilt, David},
  journal={Reviews of Modern Physics},
  volume={84},
  number={4},
  pages={1419--1475},
  year={2012},
  publisher={APS}
}

@article{krishnamoorthy2021dielectric,
  title={Dielectric constant of liquid water determined with neural network quantum molecular dynamics},
  author={Krishnamoorthy, Aravind and Nomura, Ken-ichi and Baradwaj, Nitish and Shimamura, Kohei and Rajak, Pankaj and Mishra, Ankit and Fukushima, Shogo and Shimojo, Fuyuki and Kalia, Rajiv and Nakano, Aiichiro and others},
  journal={Physical Review Letters},
  volume={126},
  number={21},
  pages={216403},
  year={2021},
  publisher={APS}
}

@article{zhang2020deep,
  title={Deep neural network for the dielectric response of insulators},
  author={Zhang, Linfeng and Chen, Mohan and Wu, Xifan and Wang, Han and E, Weinan and Car, Roberto},
  journal={Physical Review B},
  volume={102},
  number={4},
  pages={041121},
  year={2020},
  publisher={APS}
}

@article{ghosez1998dynamical,
  title={Dynamical atomic charges: The case of AB O 3 compounds},
  author={Ghosez, Ph and Michenaud, J-P and Gonze, Xavier},
  journal={Physical Review B},
  volume={58},
  number={10},
  pages={6224},
  year={1998},
  publisher={APS}
}

@article{kutana2025representing,
  title={Representing Born effective charges with equivariant graph convolutional neural networks},
  author={Kutana, Alex and Shimizu, Koji and Watanabe, Satoshi and Asahi, Ryoji},
  journal={Scientific Reports},
  volume={15},
  number={1},
  pages={16719},
  year={2025},
  publisher={Nature Publishing Group UK London}
}

@article{schmiedmayer2024derivative,
  title={Derivative learning of tensorial quantities—Predicting finite temperature infrared spectra from first principles},
  author={Schmiedmayer, Bernhard and Kresse, Georg},
  journal={The Journal of Chemical Physics},
  volume={161},
  number={8},
  year={2024},
  publisher={AIP Publishing}
}

@article{wolfsberg1952spectra,
  title={The spectra and electronic structure of the tetrahedral ions MnO4-, CrO4--, and ClO4-},
  author={Wolfsberg, MAX and Helmholz, Lindsay},
  journal={The Journal of Chemical Physics},
  volume={20},
  number={5},
  pages={837--843},
  year={1952},
  publisher={American Institute of Physics}
}

@misc{mlist,
  title = {List of organic Solvents with Information about Hansen Solubility Parameter, Solvent-Properties, Hazardousness and Cost-Analysis},
  howpublished = {\url{https://data.mendeley.com/datasets/b4dmjzk8w6/1}},
}

@article{thompson2022lammps,
  title={LAMMPS-a flexible simulation tool for particle-based materials modeling at the atomic, meso, and continuum scales},
  author={Thompson, Aidan P and Aktulga, H Metin and Berger, Richard and Bolintineanu, Dan S and Brown, W Michael and Crozier, Paul S and In't Veld, Pieter J and Kohlmeyer, Axel and Moore, Stan G and Nguyen, Trung Dac and others},
  journal={Computer physics communications},
  volume={271},
  pages={108171},
  year={2022},
  publisher={Elsevier}
}

@article{abraham2015gromacs,
  title={GROMACS: High performance molecular simulations through multi-level parallelism from laptops to supercomputers},
  author={Abraham, Mark James and Murtola, Teemu and Schulz, Roland and P{\'a}ll, Szil{\'a}rd and Smith, Jeremy C and Hess, Berk and Lindahl, Erik},
  journal={SoftwareX},
  volume={1},
  pages={19--25},
  year={2015},
  publisher={Elsevier}
}

@article{he2020fast,
  title={A fast and high-quality charge model for the next generation general AMBER force field},
  author={He, Xibing and Man, Viet H and Yang, Wei and Lee, Tai-Sung and Wang, Junmei},
  journal={The Journal of chemical physics},
  volume={153},
  number={11},
  year={2020},
  publisher={AIP Publishing}
}

@article{jakalian2002fast,
  title={Fast, efficient generation of high-quality atomic charges. AM1-BCC model: II. Parameterization and validation},
  author={Jakalian, Araz and Jack, David B and Bayly, Christopher I},
  journal={Journal of computational chemistry},
  volume={23},
  number={16},
  pages={1623--1641},
  year={2002},
  publisher={Wiley Online Library}
}

@misc{cpmd,
  title = {Car-Parrinello Molecular Dynamics},
  howpublished = {\url{https://github.com/CPMD-code}},
}

@misc{tensorflow,
title={TensorFlow},
howpublished={\url{https://www.tensorflow.org/}},
}

@article{kingma2014adam,
  title={Adam: A method for stochastic optimization},
  author={Kingma, Diederik P and Ba, Jimmy},
  journal={arXiv preprint arXiv:1412.6980},
  year={2014}
}

@article{goedecker1996separable,
  title={Separable dual-space Gaussian pseudopotentials},
  author={Goedecker, Stefan and Teter, Michael and Hutter, J{\"u}rg},
  journal={Physical Review B},
  volume={54},
  number={3},
  pages={1703},
  year={1996},
  publisher={APS}
}

@article{lee1988development,
  title={Development of the Colle-Salvetti correlation-energy formula into a functional of the electron density},
  author={Lee, Chengteh and Yang, Weitao and Parr, Robert G},
  journal={Physical review B},
  volume={37},
  number={2},
  pages={785},
  year={1988},
  publisher={APS}
}

@article{becke1988density,
  title={Density-functional exchange-energy approximation with correct asymptotic behavior},
  author={Becke, Axel D},
  journal={Physical review A},
  volume={38},
  number={6},
  pages={3098},
  year={1988},
  publisher={APS}
}

@article{kubo1957statistical,
  title={Statistical-mechanical theory of irreversible processes. I. General theory and simple applications to magnetic and conduction problems},
  author={Kubo, Ryogo},
  journal={Journal of the physical society of Japan},
  volume={12},
  number={6},
  pages={570--586},
  year={1957},
  publisher={The Physical Society of Japan}
}

@article{ro2000temperature,
  title={Temperature dependence of the dielectric function of C 6 H 6 (l) and C 6 H 5 CH 3 (l) measured with THz spectroscopy},
  author={R{\o}nne, Cecilie and Jensby, Kasper and Loughnane, Brian J and Fourkas, John and Nielsen, O Faurskov and Keiding, So/ren R},
  journal={The Journal of Chemical Physics},
  volume={113},
  number={9},
  pages={3749--3756},
  year={2000},
  publisher={American Institute of Physics}
}

@inproceedings{kakutani2021material,
  title={Material design and high frequency characterization of novel ultra-low loss dielectric material for 5G and 6G applications},
  author={Kakutani, Takenori and Suzuki, Yuya and Koh, Meiten and Sekiguchi, Shoya and Matsumura, Satoko and Oki, Kota and Mishima, Shoko and Ishikawa, Nobuhiro and Ogata, Toshiyuki and Erdogan, Serhat and others},
  booktitle={2021 IEEE 71st Electronic Components and Technology Conference (ECTC)},
  pages={538--543},
  year={2021},
  organization={IEEE}
}

@article{zhai2022terahertz,
  title={Terahertz dielectric characterization of low-loss thermoplastics for 6G applications},
  author={Zhai, Min and Locquet, Alexandre and Citrin, David S},
  journal={International Journal of Wireless Information Networks},
  volume={29},
  number={3},
  pages={269--274},
  year={2022},
  publisher={Springer}
}

@article{hohenberg1964inhomogeneous,
  title={Inhomogeneous electron gas},
  author={Hohenberg, Pierre and Kohn, Walter},
  journal={Physical Review},
  volume={136},
  number={3B},
  pages={B864},
  year={1964},
  publisher={APS}
}

@article{kohn1965self,
  title={Self-consistent equations including exchange and correlation effects},
  author={Kohn, Walter and Sham, Lu Jeu},
  journal={Physical Review},
  volume={140},
  number={4A},
  pages={A1133},
  year={1965},
  publisher={APS}
}

\appendix 
\section{Computational methods for EAD}
\subsection{Atomic basis orbitals}\label{seca:orb}
The atomic basis orbitals are $sp^n$ Gaussian hybridized orbitals with correct orientation for bonding (while not discussed here, $d$ or $f$ orbitals can be constructed with the same method):
\[\chi(\alpha,\beta,\hat{\bm p};\bm r)=\frac1{\mathcal N}(\alpha+\beta \hat{\bm p}\cdot\bm r)\exp(-a r^2),\]
where $\alpha$ and $\beta$ are the proportion of $s$ and $p$ orbitals, respectively; $\hat{\bm p}$ is the direction of $p$ orbital; $\mathcal N$ is the normalization factor; and finally the exponent $a$ is determined by fitting the Slater-type basis set and fixed for a given atom type.

For hydrogen atoms, there is only one $s$ orbital ($\alpha=1$, $\beta=0$). 
For any other atom belonging to $p$ blocks, 4 orbitals are constructed by considering its chemical bonding using \ALGO{alg:orb}.
For $j\le \textsub N{neigh}$ ($\textsub N{neigh}$ is the number of bonding neighbors), $\chi_j$ constructed this way approximately points to the $j$-th neighbor and can form a $\sigma$ bond; all other orbitals are for used for $\pi$ bonds (or lone pairs). 

\begin{algorithm}[H]
\caption{Algorithm to construct $sp^n$ basis orbitals for $p$-blocks atoms}\label{alg:orb}
\begin{algorithmic}[1]
    \Require{$\textsub N{neigh}\le 4$ the number of its bonding neighbors with relative unit displacement $\hat{\bm r}_i$ (pointing towards neighbors, $i\in\{0,\cdots,\textsub N{neigh}-1\}$)}
    \Ensure{Orbitals parameters $\alpha_j$ and $\beta_j\hat{\bm p}_j$, $j\in\{0,\cdots,3\}$}
    \State{Let $\phi_i(\bm r)=\left\{\frac12\exp(-r^2),x\exp(-r^2),y\exp(-r^2),z\exp(-r^2)\right\}$} \Comment{$s$-, $p_x$-, $p_y$- and $p_z$-like functions, respectively}
    \State{Construct $4\times\textsub N{neigh}$ matrix $A_{ij}=\phi_i(\hat{\bm r}_j)$}
    \State{Compute the full singular value decomposition (SVD) $A=USV^\top$}
    \State{$F\leftarrow U (V^\top \oplus O)$, where $O$ is an arbitrary $(4-\textsub N{neigh})\times(4-\textsub N{neigh})$ orthogonal matrix } \Comment{``$\oplus$'' denotes direct sum}
    \State{$\alpha_j\leftarrow F_{0j}$ and $\beta_j\hat{\bm p}_j\leftarrow F_{1:,j}$} \Comment{``:'' denotes array slicing}
\end{algorithmic}
\end{algorithm}

\subsection{$X^{-1}$ matrix}\label{seca:mat}

The iterative algorithm to compute the scaling parameter in subsection \ref{sec:mao} is given below. Evidently, when the iteration converges, the first condition for $X^{-1}$ (charge neutrality) will be satisfied.
\begin{algorithm}[H]
\caption{Algorithm to compute $X^{-1}$ matrix}\label{alg:xm}
\begin{algorithmic}[1]
    \Require{Subsystem formed by $N$ atoms, $\textsub Mc$ basis orbitals and $2\textsub Qc$ electrons}
    \Require{Atom $I\in\{0,1,\cdots,N-1\}$ in the subsystem provides $2q_I$ electrons and $m_I$ basis orbitals (therefore $\sum_I q_I=\textsub Qc$ and $\sum_I m_I=\textsub Mc$)}
    \Require{Molecular orbitals $\Psi_j$ of the subsystem are expressed by atomic basis orbitals $\chi_{(Ii)}$ as $\Psi_j=\sum_{Ii} \chi_{(Ii)}X_{(Ii)j}$}
    \Ensure{The MO-to-MAO transfer matrix $X^{-1}$}
    \State{Introduce scaling parameter $\ee^{\lambda_I}$ for each atom $I$, initialized to $\lambda_I\leftarrow 0$} \Comment{$\ee^{\lambda_I}$ is used in place of $\lambda_I$ in the main text}
    \Repeat
    \State{Compute the reduced SVD of matrix $X_{(Ii)j}\ee^{\lambda_I}=USV^\top$}
    \State{$X^{-1}\leftarrow VU^\top$}
    \State{$\Delta_I\leftarrow q_I-\sum_{ji}\left(X^{-1}_{j(Ii)}\right)^2$}
    \State{$\lambda_I\leftarrow\lambda_I+\eta \Delta_I$} \Comment{$\eta>0$ is the update step size}
    \Until{All $|\Delta_I|$ are sufficiently small}
\end{algorithmic}
\end{algorithm}

To discuss the convergence properties of this iteration, we define positive semidefinite matrices $A_{Ikl}=\sum_i X_{(Ii)k}X_{(Ii)l}$ and function
\[F(\{\lambda_I\})=\log \det \left(\sum_I \ee^{\lambda_I}A_I\right)-\sum_I q_I\lambda_I.\]
It can be verified that $F$ is convex and $-\frac{\partial F}{\partial \lambda_I}=\Delta_I$ above,
therefore, the fixed point of the iteration is unique. However, when $\frac{q_I}{m_I}$ are unity for some atoms (e.g., in the case of $p$-$\pi$ conjugacy as N atom in peptide bond), 
the iteration has no fixed point.
The $\lambda_I$ of these atoms are in fact $\infty$ and require special treatment, given by the slightly modified \ALGO{alg:xm2} below. While mathematically it is still possible to construct pathological cases where the iteration does not converge, in practice, we have numerically verified that this modified algorithm always works.

\begin{algorithm}[H]
\caption{Modified algorithm to compute $X^{-1}$ matrix}\label{alg:xm2}
\begin{algorithmic}[1]
    \Require{(Same as \ALGO{alg:xm})}
    \Ensure{The MO-to-MAO transfer matrix $X^{-1}$ (Same as \ALGO{alg:xm})}
    \State{Sort the atoms such that $\frac{q_I}{m_I}\le\frac{q_{I+1}}{m_{I+1}}$}
    \State{Let constant $P=1+(\text{max value of $I$ such that $\frac{q_I}{m_I}<1$})$ and $R=\sum_{I\ge P} m_I$} \Comment{All indices range from 0 to $\text{length}-1$}
    \State{Introduce scaling parameter $\ee^{\lambda_I}$ \emph{only for $I<P$}, initialized to $\lambda_I\leftarrow 0$}
    \Repeat
    \State{Compute matrix $Y$ whose elements are $Y_{(Ii),j}=\begin{cases}
  X_{(Ii)j}\ee^{\lambda_I}  & I< P\\
  X_{(Ii)j}               & I\ge P
\end{cases}$}
    \State{Compute the full SVD of submatrix $Y_{(P:,:),:}=\tilde U \tilde S \tilde V^\top$} \Comment{Note that $Y_{(P:,i)j}$ actually does not change in the loop}
    \State{Compute the reduced SVD of submatrix $Y_{(:P,:),:}\tilde V_{:,R:} =USV^\top$}
    \State{$X^{-1}\leftarrow \left[ (\tilde V_{:,R:})^\top VU^\top\middle|\tilde V_{:,:R}\tilde U^\top\right]$} \Comment{``$|$'' denotes concatenation along rows}
    \State{$\Delta_I\leftarrow q_I-\sum_{ji}\left(X^{-1}_{j(Ii)}\right)^2$} \Comment{Note that $\Delta_{P:}$ all vanish automatically}
    \State{$\lambda_I\leftarrow\lambda_I+\eta \Delta_I$}
    \Until{All $|\Delta_I|$ are sufficiently small}
\end{algorithmic}
\end{algorithm}

\section{Machine learning models}\label{seca:nn}
\begin{figure}
\includegraphics[scale=.6]{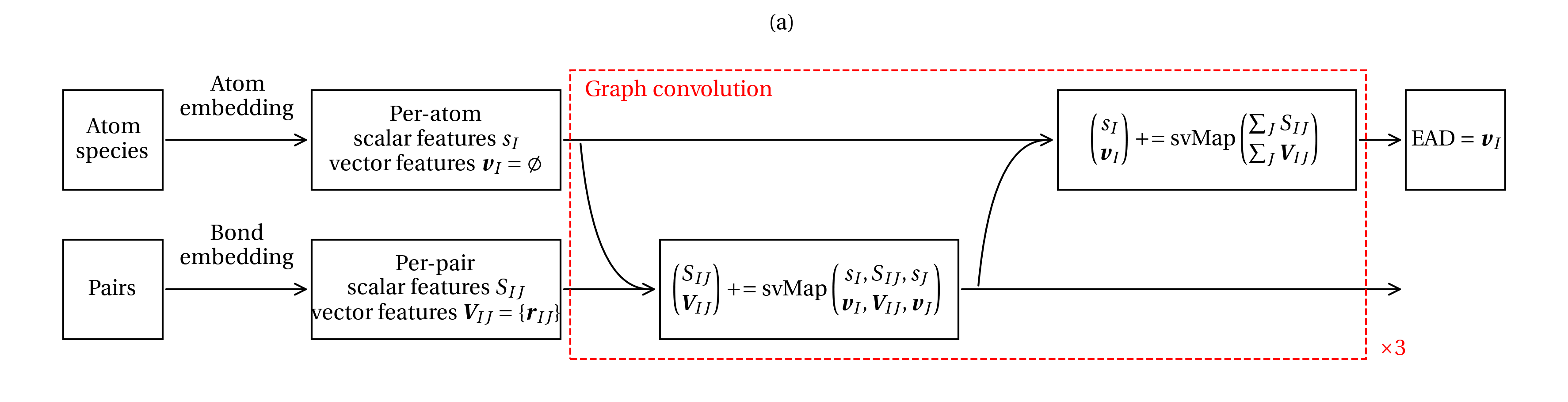}
\includegraphics[scale=.6]{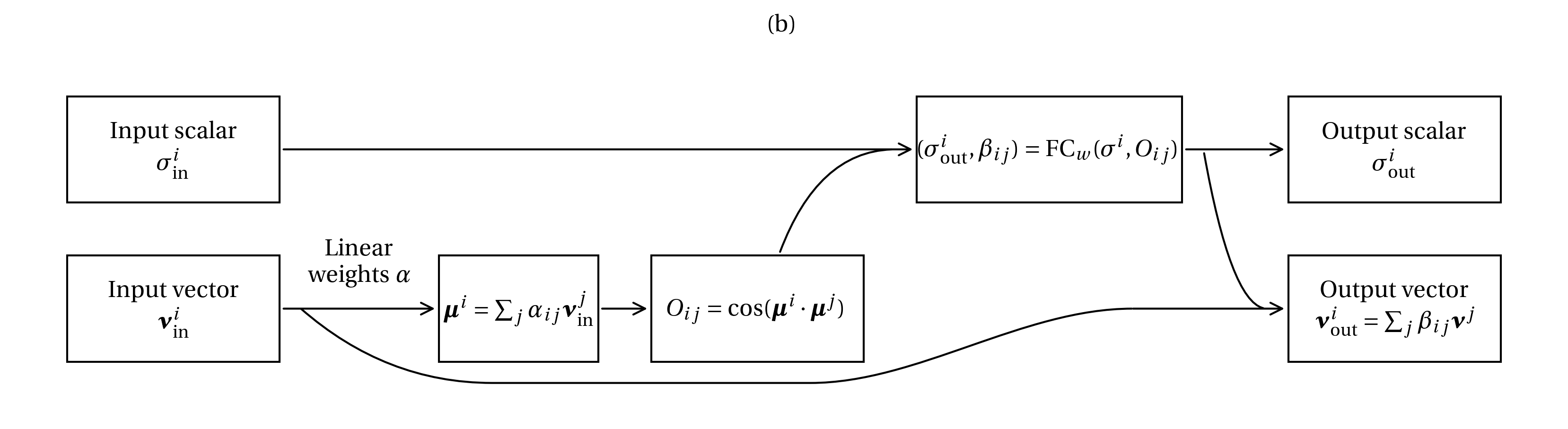}
\caption{(a) The architecture of the graph neural network used to predict the EAD. The red dashed box represents the graph convolution and is applied 3 times with independent weights. (b) The architecture of the $\svMap$ layer which is the building block of the graph convolution layer. The linear weights $\alpha$ and the weights of fully connected layer (FC) $w$ are trainable parameters.}\label{fg:nnstr}
\end{figure}

We have constructed a graph neural network model inspired by CHGNet \cite{deng2023chgnet} to predict the EAD (and also the total dipole moment) from atomic species and positions, as shown in \FIG{fg:nnstr}.
Each atom $I$ carries a set of scalar features $s_I$, initialized using an atom embedding layer, and vector features $\bm v_I\equiv (v_{Ix},v_{Iy},v_{Iz})$ initialized as empty set.
Each atomic pair $IJ$ within a cutoff of \SI{6}{\angstrom} also carries a set of scalar features $S_{IJ}$ initialized by Fourier encoding of bond lengths \cite{deng2023chgnet}, and vector features $\bm V_{IJ}$ initialized as the relative displacement between atom $I$ and $J$.
We then repeatedly update the per-atom and per-pair features using graph convolution [the rad box in \FIG{fg:nnstr} (a)] built upon the rotationally covariant mapping layers referred to as ``$\svMap$'' (explained below).
The graph convolution is applied for a total of 3 times with independent weights each time, and the final vector features on each atom are identified as the output EAD.

The $\svMap$ layer has similar to the neural network model in Ref \cite{zhang2020deep} and
takes a set of scalars $\sigma^i_{\text{in}}$ and vectors $\bm\nu^i_{\text{in}}$ as inputs (either can be an empty set) and outputs another set of scalars $\sigma^i_{\text{out}}$ and vectors $\bm\nu^i_{\text{out}}$ (either can be an empty set).
The rotational covariance indicates that if we keep all input scalars unmodified and apply an arbitrary rotation operation to all input vectors, the output scalars will not change but the vectors will also be rotated the same way.
As shown in \FIG{fg:nnstr} (b), the key to achieve this covariance is that the neural network never directly processes the vector inputs but only the intermediate scalar values computed from vectors, such as the cosine between the vectors, and output vectors are the linear combination with coefficients output by the neural network.

\end{document}